\documentclass[conference,9pt]{IEEEtran}

\IEEEoverridecommandlockouts
\usepackage{cite}
\usepackage{amsmath,amssymb,amsfonts}
\usepackage{algorithmic}
\usepackage{graphicx}
\usepackage{textcomp}
\usepackage[table]{xcolor}

\usepackage{xurl}

\usepackage{graphicx}
\usepackage{pifont}
\usepackage{newunicodechar}
\newcommand{\circlednum}[1]{\scalebox{1.25}{\ding{#1}}}
\newunicodechar{❶}{\circlednum{202}}
\newunicodechar{❷}{\circlednum{203}}
\newunicodechar{❸}{\circlednum{204}}
\newunicodechar{❹}{\circlednum{205}}
\newunicodechar{❺}{\circlednum{206}}

\usepackage{algorithm}
\usepackage{array}
\definecolor{hlLane}{HTML}{DCEBF7}
\definecolor{hlByte}{HTML}{DDF0E7}
\definecolor{hlAcc}{HTML}{FFF1B8}
\definecolor{hlRnd}{HTML}{FDE3CF}
\definecolor{hlAct}{HTML}{EADDF3}
\definecolor{hlMrg}{HTML}{F6D3E8}
\newcommand{\kw}[1]{\textbf{#1}}
\newcommand{\cm}[1]{{\color{black!55}$\triangleright$\,#1}}
\newcommand{\ttl}[1]{\multicolumn{2}{@{}l}{\textbf{#1}}}
\newcommand{\inp}[1]{\multicolumn{2}{@{}>{\raggedright\arraybackslash}p{0.985\linewidth}@{}}{\textbf{Input:} #1}}
\newcommand{\outp}[1]{\multicolumn{2}{@{}>{\raggedright\arraybackslash}p{0.985\linewidth}@{}}{\textbf{Output:} #1}}
\newcommand{\cd}[1]{\texttt{#1}}

\usepackage{booktabs}
\usepackage[hidelinks]{hyperref}

\def\BibTeX{{\rm B\kern-.05em{\sc i\kern-.025em b}\kern-.08em
    T\kern-.1667em\lower.7ex\hbox{E}\kern-.125emX}}
\begin{document}

\title{IsoMoE: Placement-Invariant Hybrid CPU-GPU MoE Inference with Lossless 12-bit Experts\\
}

% \and
% \and
% \and
% \and
% \and

\author{%
\IEEEauthorblockN{Hang Gu\textsuperscript{1}, Teng Wang\textsuperscript{2}, Zhendong Zheng\textsuperscript{1},
Qianyu Cheng\textsuperscript{1}, Wenqi Lou\textsuperscript{2}, Lei Gong\textsuperscript{1}, Xi Li\textsuperscript{1},
Xuehai Zhou\textsuperscript{1,2}}
\IEEEauthorblockA{\textsuperscript{1}University of Science and Technology of China, Hefei, China\\
\textsuperscript{2}Suzhou Institute for Advanced Research, University of Science and Technology of China, Suzhou,
China\\
\{hanggu, zzd1411, qycheng\}@mail.ustc.edu.cn, \{wangt635, louwenqi, leigong0203, llxx, xhzhou\}@ustc.edu.cn}}

\maketitle

\begin{abstract}
Mixture-of-Experts (MoE) models scale model capacity at low computational cost. Hybrid CPU-GPU inference offloads some experts to the CPU, enabling a single GPU to run MoE models larger than its memory. Expert placement changes with free GPU memory and should only affect speed. However, the greedy decoding outputs of existing hybrid engines change with placement. This is because the two devices use different arithmetic to compute experts. In addition, decoding is limited by CPU memory bandwidth. Lossless compression reduces memory traffic, but the CPU decompresses entropy-coded formats too slowly to keep up with memory reads. To address these challenges, we propose IsoMoE, an efficient hybrid MoE inference system that guarantees that its outputs do not change with expert placement. IsoMoE introduces (i) a GPU expert kernel that bitwise replays the arithmetic of the CPU expert kernel, (ii) a memory layout that allows the CPU to efficiently compute lossless 12-bit experts using AVX-512 dot-product instructions, and (iii) a dual micro-batch pipeline that overlaps CPU and GPU computation without changing the outputs. Experimental results on three MoE models show that IsoMoE produces bitwise identical outputs across different expert placements. It is also faster than the fastest baseline in every setting, achieving an average decoding speedup of 1.26\(\times\) and a maximum speedup of 1.54\(\times\). Our source code is publicly available at \url{https://anonymous.4open.science/r/isomoe-artifact-9C21/}.
\end{abstract}


% ----------------------------------------第一章----------------------------------------
\section{Introduction}

Mixture-of-Experts (MoE) models scale model capacity at low computational cost \cite{jiang2024mixtralexperts,deepseekai2024deepseekv2strongeconomicalefficient,yang2025qwen3technicalreport}. However, their experts often do not fit in the memory of a single GPU. For example, the experts of Qwen3-235B-A22B \cite{yang2025qwen3technicalreport} occupy about 454 GB in BF16, and even a GPU with 80 GB of memory can hold less than one fifth of them. Therefore, hybrid CPU-GPU inference computes some experts on the GPU and the rest on the CPU, enabling a single GPU to run MoE models larger than its memory \cite{kamahori2025fiddler,chen2025ktransformers,gerganov2023llamacpp,son2026cox,yang2026freetoken,zhong2025hybrimoe}. We refer to this allocation of experts between the two devices as expert placement. Expert placement changes with the expert GPU memory budget. By default, llama.cpp~\cite{gerganov2023llamacpp} and FreeToken~\cite{yang2026freetoken} set this budget based on the free GPU memory at startup. The GPU is not necessarily dedicated to inference, so its free memory may vary across launches~\cite{yang2026freetoken}. FreeToken and HybriMoE~\cite{zhong2025hybrimoe} also change the placement during inference.

Changing expert placement should only affect inference speed and should not change the outputs. Otherwise, the output of the same command depends on what else occupies GPU memory at startup. Batch-invariant kernels~\cite{he2025nondeterminism} can make outputs independent of batch size, but they cannot make outputs independent of expert placement. Consequently, researchers evaluating large MoE models on a single GPU cannot tell whether an accuracy improvement comes from their method or from changes in free GPU memory~\cite{yuan2026understanding}. Engine developers also cannot use the outputs of all-CPU runs to verify new caching or swap-in strategies.

However, the greedy decoding outputs of existing hybrid engines change with expert placement. When Qwen3-30B-A3B solves 56 math problems~\cite{h4aime2024,zhang2025aime25,lightman2024let}, only 5 problems have unchanged outputs after llama.cpp moves the experts of 12 layers to the GPU. Halving the FreeToken expert cache changes all 56 outputs. If another process occupies 4 GiB of GPU memory at startup, llama.cpp keeps 4.1 GiB more experts on the CPU, and only 3 out of 16 math problems have unchanged outputs. The reason is that the CPU and GPU use different arithmetic to compute experts. When the routing probabilities of two experts are almost equal, a subtle difference in rounding can swap the selected expert. Caching~\cite{yang2026freetoken,zhong2025hybrimoe} and compression~\cite{tan2026approaching,yang2026zipmoe} both improve speed by placing more experts on the GPU, so they also change the outputs in existing engines. Therefore, users of existing hybrid engines must choose between speed and unchanged outputs.

In addition, decoding is limited by CPU memory bandwidth. To guarantee unchanged outputs, existing engines must keep all experts on the CPU, where CPU expert computation accounts for about 82\% of each decode step of Qwen3-30B-A3B. To accelerate decoding, the CPU must read fewer expert bytes. Common approaches are quantization and lossless compression \cite{tan2026approaching,yang2026zipmoe}. However, both approaches have shortcomings. Quantization changes the weights, and the answers to some problems change accordingly \cite{dutta2024accuracy}. Lossless compression preserves the released weights, but the CPU can only decompress existing entropy-coded formats \cite{zhang202570,yang2026zipmoe,tan2026approaching} symbol by symbol, too slowly to keep up with memory reads. With these formats, CPU expert computation is only 0.59 to 1.07 times the speed of BF16, far below the expected speedup of about 1.5 from reading fewer bytes. Fixed-length 12-bit codewords can be decompressed in parallel, but their decompression kernels are designed for the GPU \cite{liberman2026nyoomfloat12,koritala2026glyd}.

% 考虑了下还是把这个图删掉。首先这个图本来就没什么用，凑空间用的。另外，论文的动机讲的还不算太好，删掉这个图之后给introduction加点论述
% \begin{figure}[t]
% \centering
% \includegraphics[width=\columnwidth]{figures/fig1_teaser.pdf}
% \caption{Effect of expert placement on outputs. (a) Existing hybrid engines use different arithmetic on the CPU and GPU, so the outputs change. (b) The GPU of IsoMoE follows the CPU in bytes and arithmetic.}
% \label{fig:teaser}
% \end{figure}

To address these challenges, we propose IsoMoE, an efficient placement-invariant hybrid MoE inference system. For the same hardware, software, and batch of requests, IsoMoE's outputs are bitwise identical regardless of which experts are in GPU memory and when they are swapped in. These outputs also equal the outputs obtained when all BF16 experts are on the CPU. IsoMoE introduces three techniques. (i) Asymmetric CPU-GPU replay. CPU expert computation is the bottleneck, while the GPU is often idle. Therefore, IsoMoE keeps the arithmetic of KTransformers \cite{chen2025ktransformers} in the CPU expert kernel and lets the GPU expert kernel replay it bitwise. (ii) A memory layout that allows the CPU to efficiently compute lossless 12-bit experts. We call it CPU order. IsoMoE adopts fixed-length 12-bit codewords \cite{liberman2026nyoomfloat12,koritala2026glyd} and arranges their bytes in the order the CPU expert kernel feeds weights into dot-product instructions. The CPU can then decompress a whole vector in registers at once, and CPU expert computation reaches 1.35 times the speed of BF16, as expected from reading about one quarter fewer bytes. The 12-bit experts are smaller, so the same GPU memory can hold more experts. Placing more experts on the GPU changes the placement but does not change IsoMoE's outputs. (iii) An exact dual micro-batch pipeline. It follows the schedule of DeepSeek-V3 \cite{liu2024deepseek} to overlap CPU and GPU computation and computes each micro-batch using the whole-batch shape, so overlapping computation does not change the outputs.

In summary, our contributions are as follows:

\begin{itemize}
\item We analyze why the outputs of existing hybrid engines change with expert placement and why the CPU decompresses existing entropy-coded formats too slowly.
\item We propose an asymmetric CPU-GPU replay mechanism that allows each expert in a layer to be placed individually on the CPU or GPU without changing the outputs.
\item We propose a memory layout called CPU order that enables CPU expert computation to achieve the expected speedup from reading fewer bytes.
\item We implement IsoMoE on top of KTransformers \cite{chen2025ktransformers}. Experimental results show that on three MoE models, IsoMoE produces bitwise identical outputs across different expert placements and is faster than the fastest baseline in every evaluated setting, by 1.26$\times$ on average and up to 1.54$\times$ in decoding throughput.
\end{itemize}

% ----------------------------------------第二章----------------------------------------
\section{Background}
\label{sec:background}

\subsection{MoE Models and Expert Offloading}
MoE models replace the feed-forward network in a Transformer layer with a set of experts, and each token activates only a few of these experts \cite{jiang2024mixtralexperts,deepseekai2024deepseekv2strongeconomicalefficient,yang2025qwen3technicalreport}. For an input token $x$, each MoE layer selects the $k$ experts $\mathcal{S}(x)$ with the highest routing probabilities and outputs their weighted sum ${y=\sum_{i\in \mathcal{S}(x)} g_i(x)E_i(x)}$, where $E_i(x)$ is the output of expert $i$ and $g_i(x)$ is its routing weight. The experts of Qwen3-30B-A3B occupy 58 GB in BF16, more than the 48 GB memory of an NVIDIA RTX A6000 GPU \cite{nvidia2022a6000}, which also has to hold the KV cache.
To address this problem, many works use expert offloading, which keeps expert weights in host memory. One class of methods computes experts only on the GPU and uses prefetching and caching to reduce the overhead of moving experts over PCIe \cite{hwang2024pre,he2024expertflow}. Another class of methods lets the CPU compute the experts that are not in GPU memory. Fiddler \cite{kamahori2025fiddler}, KTransformers \cite{chen2025ktransformers}, llama.cpp \cite{gerganov2023llamacpp}, and CoX-MoE \cite{son2026cox} decide which experts to keep in GPU memory before inference, while FreeToken \cite{yang2026freetoken} and HybriMoE \cite{zhong2025hybrimoe} replace them during inference.

\subsection{Deterministic LLM Inference}
Execution configurations such as batch size and the number and type of GPUs should only affect inference speed, but they may change the output even with greedy decoding \cite{yuan2026understanding,zhang2026deterministic}. Such output changes hinder evaluation~\cite{yuan2026understanding} and auditing~\cite{badash2026hawkeye}, and also make reinforcement learning unstable, especially with MoE routing~\cite{zhang2026deterministic,ma2026stabilizing}.
Execution configurations can change the output mainly because floating-point addition is not associative and different hardware rounds intermediate results differently. To achieve deterministic LLM inference, one line of work makes the output independent of batch size \cite{he2025nondeterminism,sglang2025towards,gond2026llm}. Batch-invariant kernels use the same reduction order regardless of batch size. They cannot select the fastest reduction strategy for each batch size, so inference slows down \cite{he2025nondeterminism}. For example, when deterministic mode is enabled in SGLang \cite{sglang2025towards}, the throughput of Llama-3 models on H100 GPUs drops by up to 52\% \cite{gond2026llm}. Another line of work makes different hardware produce bitwise identical results by running the same kernels on all hardware \cite{cooper2026accelerating,arun2025verde} or by reproducing GPU Tensor Core matrix multiplication on the CPU \cite{badash2026hawkeye}. Neither line of work addresses expert placement in hybrid inference.

\subsection{Lossless Weight Compression}
The decode phase is memory-bandwidth-bound on both the GPU and the CPU \cite{liberman2026nyoomfloat12,gerogiannis2025deca}. Compressing expert weights reduces the bytes read and allows GPU memory to hold more experts \cite{tan2026approaching,yang2026zipmoe}. Quantization saves more space but changes the weights. Even if overall accuracy is almost unchanged, answers flip on some problems \cite{dutta2024accuracy}. Lossless compression keeps the decompressed weights bitwise identical to the released BF16 weights and shrinks them to about 70\% of their size. This is because the 8-bit exponent carries only about 2.6 bits of information on average \cite{zhang202570,yang2026zipmoe}.
Most existing lossless formats are decompressed on the GPU. Some use Huffman codes or rANS for entropy coding, and their bit rates are close to the entropy lower bound \cite{zhang202570,tan2026approaching}. Fixed-length formats such as NyoomFloat12 \cite{liberman2026nyoomfloat12} and Glyd \cite{koritala2026glyd} use 12-bit codewords. Decompressing quantized and sparse weights with vector instructions on the CPU cannot keep up with memory reads when bandwidth per core is high or each weight requires many instructions \cite{gerogiannis2025deca}.

% ----------------------------------------第三章----------------------------------------
\section{Motivation}
\label{sec:motivation}

\begin{figure}[t]
\centering
\includegraphics[width=0.9\columnwidth]{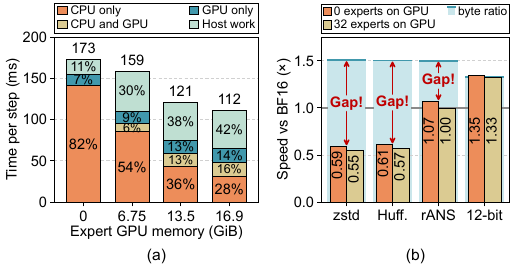}
\caption{CPU expert path of Qwen3-30B-A3B at batch size 8. (a) Decode step breakdown with 0 to 16.9~GiB of BF16 experts in GPU memory. Host work is the main thread's scheduling, data copies, and kernel launches. (b) Speed of CPU expert computation with lossless formats, relative to BF16.}
\label{fig:cpu_path}
\end{figure}

\begin{figure}[t]
\centering
\includegraphics[width=0.9\columnwidth]{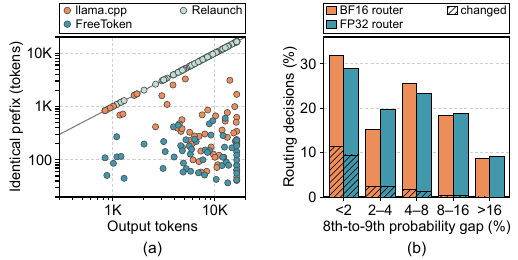}
\caption{Effect of changing only the placement on Qwen3-30B-A3B. (a) Identical prefix per problem, with outputs capped at 16K tokens. (b) Routing decisions by gap.}
\label{fig:placement}
\end{figure}

As shown in Fig.~\ref{fig:cpu_path}(a), when all BF16 experts are on the CPU, CPU expert computation accounts for 82\% of each decode step. Its memory read rate is about 127~GB/s, which reaches the maximum measured memory read bandwidth. Even with 13.5~GiB of BF16 experts in GPU memory, CPU expert computation still takes 49\% of each step. Therefore, to make decoding faster while keeping outputs unchanged, the CPU must read fewer expert bytes. We can achieve this by computing more experts on the GPU or by shrinking each expert losslessly.

\textbf{Challenge 1: Placement-Dependent Outputs.} Changing expert placement should only affect speed, but it changes the greedy decoding outputs of existing hybrid engines. After llama.cpp \cite{gerganov2023llamacpp} moves the experts of 12 layers to the GPU, only 5 out of 56 problems have unchanged outputs. When the expert cache of FreeToken \cite{yang2026freetoken} is halved, all 56 outputs change. Relaunching either engine reproduces all 56 outputs, so the change comes only from placement. As shown in Fig.~\ref{fig:placement}(a), the median identical prefix length of diverged outputs is only 125 to 320 tokens. Under the default placement settings and the same launch command, both the placement and the outputs change if the free GPU memory at startup differs. For example, if an idle process occupies 4 GiB of GPU memory, llama.cpp keeps 4.1 GiB more experts on the CPU, and only 3 out of 16 math problems have unchanged outputs within the first 2,048 tokens. FreeToken caches 3.8 GiB fewer experts, and all 16 outputs change. When either engine is relaunched without the idle process, all 16 outputs are reproduced. The outputs change with placement because the CPU and GPU use different accumulation orders, rounding, and activation implementations when computing the same expert. Top-8 routing amplifies this difference and can swap the 8th selected expert with the 9th unselected expert when their probabilities are almost equal. In the PyTorch implementation of the model, we introduce rounding-level differences only to the 87 experts per layer not in GPU memory. As shown in Fig.~\ref{fig:placement}(b), with a BF16 router, 15.8\% of routing decisions select a different expert, and most of them occur when probabilities are almost equal. Even with a higher-precision FP32 router, 13.5\% of decisions still select a different expert. Users of existing hybrid engines thus have to choose between changing placement for speed and keeping outputs unchanged.

\textbf{Opportunity 1: Bitwise Replay of CPU Arithmetic.} CPU expert computation uses a fixed set of arithmetic operations. If the GPU bitwise replays these operations, the outputs no longer change with placement.

\textbf{Challenge 2: The CPU Decompression Budget.} Shrinking experts requires the CPU to decompress while reading memory. On a 48-core Intel Xeon Platinum 8488C CPU \cite{qiu2024m7i}, we compute experts with 32 threads, and each thread gets about 4 GB/s of memory bandwidth on average. The zstd used by ZipMoE \cite{yang2026zipmoe}, the Huffman code of DFloat11 \cite{zhang202570}, and rANS \cite{tan2026approaching} compress weights to about 10.7 bits, which leaves only about 1.06 cycles per weight at a runtime frequency of about 3.2 GHz. These formats are decompressed symbol by symbol. As shown in Fig.~\ref{fig:cpu_path}(b), CPU expert computation with these formats is only 0.59 to 1.07 times the speed of BF16, far below their byte ratio of about 1.5. The byte ratio is the bytes per weight of BF16 divided by those of the compressed format. It is the speedup expected without decompression overhead. The 12-bit codewords of NyoomFloat12 \cite{liberman2026nyoomfloat12} and Glyd \cite{koritala2026glyd} are fixed-length and are decompressed in parallel on the GPU.

\textbf{Opportunity 2: Fixed-Length Codewords in CPU Order.} If we arrange the bytes of 12-bit codewords in the order the CPU expert kernel feeds weights into dot-product instructions, the CPU can decompress a whole vector in registers at once and feed it directly into the dot product. As shown in Fig.~\ref{fig:cpu_path}(b), CPU expert computation then runs at 1.35 times the speed of BF16 and reaches the byte ratio.

% ----------------------------------------第四章----------------------------------------
\section{IsoMoE Design}
\label{sec:design}

\begin{algorithm*}[t]
\caption{Bitwise replay of the CPU expert kernel on the GPU. $M$ has $N$ rows and $K$ columns, $T$ holds up to 8 tokens of expert $e$ (more tokens take more passes), and lines with $t$ run for each $t \in T$. Same-color lines perform the same arithmetic.}
\label{alg:replay}
\footnotesize
\renewcommand{\arraystretch}{0.944}
\setlength{\tabcolsep}{2pt}
\begin{minipage}[t]{0.497\textwidth}
\begin{tabular}[t]{@{}r@{\hspace{4pt}}>{\raggedright\arraybackslash}p{0.93\linewidth}@{}}
\ttl{CPU expert kernel (arithmetic of KTransformers)} \\
\hline
\inp{$x_t$, routing weights $g_j$, 12-bit experts} \\
\outp{$y_t = \sum_{j \in \mathcal{S}(x_t)} g_j d_j$} \\
\hline
1 & \kw{parallel for each} task $(e, \text{64 rows})$ \kw{do} \\
\rowcolor{hlLane}2 & \quad \kw{for} matrix $M =$ gate/up \kw{do} \\
\rowcolor{hlLane}3 & \qquad \kw{for each} group $s$ of 32 rows \kw{do} \\
\rowcolor{hlLane}4 & \qquad\quad $a_0[t], a_1[t] \gets 0$ \hfill \cm{one row per FP32 lane} \\
5 & \qquad\quad \kw{for} unit $p = 0, \dots, K/2-1$ \kw{do} \\
6 & \qquad\qquad \cd{\_mm\_prefetch}(unit $p+16$) \\
7 & \qquad\qquad $\bar{x}_t \gets \cd{vpbroadcastd}(x_t[2p], x_t[2p{+}1])$ \\
\rowcolor{hlByte}8 & \qquad\qquad $w_0, w_1 \gets$ BF16 pairs of unit $p$, 16 rows each \hfill \cm{\ref{sec:cpu_order}, in zmm} \\
\rowcolor{hlAcc}9 & \qquad\qquad $a_0[t] \gets \cd{vdpbf16ps}(a_0[t], \bar{x}_t, w_0)$ \hfill \cm{odd, then even} \\
\rowcolor{hlAcc}10 & \qquad\qquad $a_1[t] \gets \cd{vdpbf16ps}(a_1[t], \bar{x}_t, w_1)$ \hfill \cm{other 16 rows} \\
\rowcolor{hlRnd}11 & \qquad\quad $z_t/u_t \gets \cd{vcvtne2ps2bf16}$ after \cd{vpermt2ps} \hfill \cm{row order} \\
\rowcolor{hlAct}12 & \quad $h_t \gets \cd{vcvtne2ps2bf16}(\mathrm{act}(z_t) \cdot u_t)$ \hfill \cm{act uses \cd{vrcp14ps}} \\
13 & $d \gets$ lines 1--11 on $h_t$ with $M =$ down \\
\rowcolor{hlMrg}14 & $y_t \gets 0$; \kw{for} $j$ in routing order: $y_t \gets \cd{fmadd}(d_j, g_j, y_t)$ \\
\rowcolor{hlMrg}15 & $y_t \gets \cd{vcvtne2ps2bf16}(y_t)$ \\
\end{tabular}
\end{minipage}\hfill
\begin{minipage}[t]{0.497\textwidth}
\begin{tabular}[t]{@{}r@{\hspace{4pt}}>{\raggedright\arraybackslash}p{0.93\linewidth}@{}}
\ttl{GPU expert kernel (replay)} \\
\hline
\inp{items $(e, T)$, $x_t$, $g_j$, 12-bit experts, ACT table, CPU $d_j$} \\
\outp{$y_t = \sum_{j \in \mathcal{S}(x_t)} g_j d_j$} \\
\hline
1 & $(e, T) \gets \mathrm{items}[\cd{blockIdx.x} / (N/32)]$ \hfill \cm{two warps per block} \\
\rowcolor{hlLane}2 & $s \gets \cd{blockIdx.x} \bmod (N/32)$; $M \gets$ gate for warp 0, up for warp 1 \\
\rowcolor{hlLane}3 & $q \gets \cd{threadIdx.x} \bmod 32$; $a[t] \gets 0$ \hfill \cm{one row per thread} \\
4 & \kw{for} $i = 0, \dots, K/16-1$ \kw{do} \hfill \cm{3 stages per warp in shared memory} \\
5 & \quad \cd{\_\_syncwarp()}; \cd{cp.async} stage $i{+}2$ \\
6 & \quad \cd{cp.async.wait\_group(2)}; \cd{\_\_syncwarp()} \hfill \cm{stage $i$ ready} \\
7 & \quad \kw{for} unit $p = 8i, \dots, 8i{+}7$ \kw{do} \\
\rowcolor{hlByte}8 & \qquad $w[2p], w[2p{+}1] \gets$ BF16 pair of row $32s{+}q$ \hfill \cm{\ref{sec:cpu_order}, in registers} \\
\rowcolor{hlAcc}9 & \qquad $a[t] \gets \cd{fma.rn.ftz}(x_t[2p{+}1], w[2p{+}1], a[t])$ \hfill \cm{odd, then even} \\
\rowcolor{hlAcc}10 & \qquad $a[t] \gets \cd{fma.rn.ftz}(x_t[2p], w[2p], a[t])$ \\
\rowcolor{hlRnd}11 & $z_t/u_t \gets \mathrm{bf16}(a[t])$ in shared memory; \cd{\_\_syncthreads()} \\
\rowcolor{hlAct}12 & \kw{if} warp 0: $h_t \gets \mathrm{bf16}(\cd{mul.rn.ftz}(\mathrm{ACT}[z_t], u_t))$ \hfill \cm{CPU act$(z)$} \\
13 & $d \gets \mathrm{bf16}$(lines 3--10 on $h_t$) with $M =$ down \\
\rowcolor{hlMrg}14 & $y_t \gets 0$; \kw{for} $j$ in routing order: $y_t \gets \cd{fma.rn.ftz}(d_j, g_j, y_t)$ \\
\rowcolor{hlMrg}15 & $y_t \gets \mathrm{bf16}(y_t)$ \hfill \cm{$d_j$ from CPU and GPU} \\
\end{tabular}
\end{minipage}
\end{algorithm*}

\subsection{Overview}
\begin{figure}[t]
\centering
\includegraphics[width=\columnwidth]{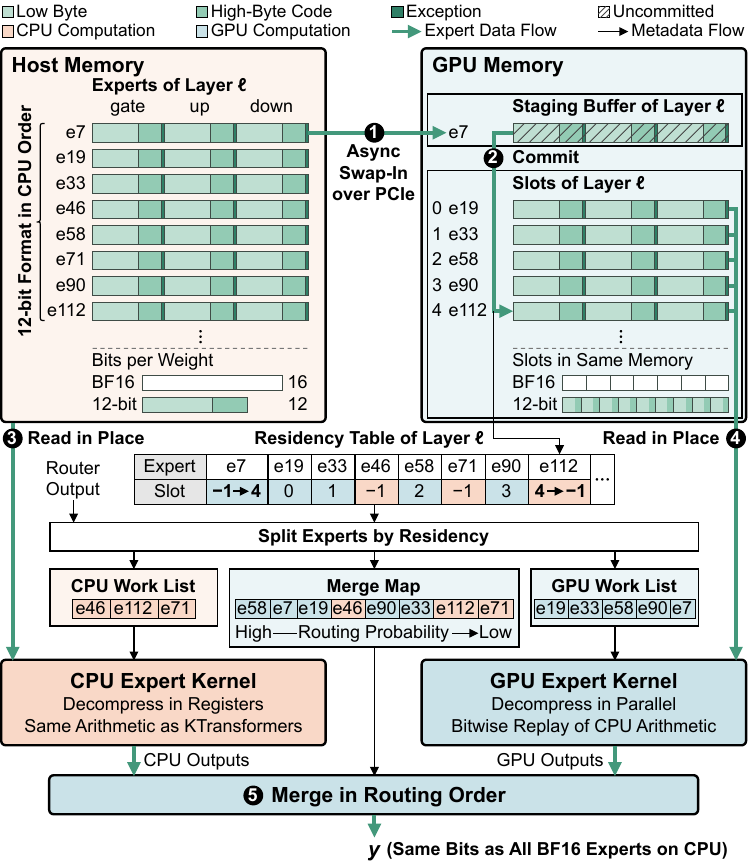}
\caption{Overview of IsoMoE. The data path of layer $\ell$ for one decode-phase token, where an asynchronous swap-in moves e7 into the slot of e112.}
\label{fig:overview}
\end{figure}

As shown in Fig.~\ref{fig:overview}, host memory holds all experts in the 12-bit format arranged in CPU order (Section~\mbox{\ref{sec:cpu_order}}), and GPU memory slots cache a subset of them. To change the placement, a swap-in ❶ first copies an expert into the staging buffer. A commit ❷ then copies it into a slot and updates the residency table. Splitting the router-selected experts by residency gives a CPU work list, a GPU work list, and a merge map. The two expert kernels in IsoMoE read their respective experts in place (❸, ❹). The CPU expert kernel retains the arithmetic of KTransformers, and the GPU expert kernel replays this arithmetic bitwise (Section~\mbox{\ref{sec:replay}}). The outputs from both sides are merged in routing order~❺, so the resulting \(y\) is bitwise identical to its value when all BF16 experts are on the CPU. IsoMoE also overlaps CPU and GPU computation with two micro-batches, and neither the micro-batches nor the timing of swap-ins changes these bits (Section~\mbox{\ref{sec:pipeline}}).

\subsection{Asymmetric Replay}
\label{sec:replay}

As shown in Fig.~\ref{fig:cpu_path}(a), even with 13.5 GiB of BF16 experts in GPU memory, only the CPU computes experts during much of each decode step while the GPU idles. Therefore, IsoMoE keeps the CPU arithmetic and lets the GPU replay it bitwise. A standard GPU kernel sums the products in a different order. Floating-point addition is not associative, so the last few bits of \(y\) may differ and change the next layer's routing (Section~\ref{sec:motivation}).

For a token \(t\), the gate and up matrices of an expert map \(x_t\) to \(z_t\) and \(u_t\), and its down matrix maps \(h_t=\mathrm{act}(z_t)\cdot u_t\) to the output \(d\). In Algorithm~\ref{alg:replay}, each output row is accumulated by one FP32 lane on the CPU (line 4) and by one thread on the GPU (line 3). Both sides read the same bytes in the 12-bit format in the same order (lines 5 to 8). On the CPU, the AVX-512 BF16 instruction \texttt{vdpbf16ps} \cite{intel2026sdm} adds the two products of a pair in turn and rounds after each addition. The odd product comes first. The GPU uses two \texttt{fma.rn.ftz} instructions in inline PTX to perform the same two steps (lines 9 and 10). They have the same rounding and denormal handling as \texttt{vdpbf16ps}, and the compiler cannot rewrite or reorder them. Both sides round \(z_t\) and \(u_t\) to BF16 in the same way (line 11).

KTransformers uses the approximate reciprocal instruction \texttt{vrcp14ps} to compute activations, and no GPU instruction can reproduce it bitwise. A BF16 \(z_t\) has only 65,536 possible values, so the GPU looks up \(\mathrm{act}(z_t)\) in a table of CPU results and multiplies it by \(u_t\) as the CPU does (line 12). The down projection repeats these dot products on \(h_t\) (line 13).

A KTransformers CPU call returns only \(y_t\). It sums the weighted outputs \(g_j d_j\) of the \(k\) experts of a token in routing order (lines 14 and 15). If both sides sum the outputs of their own experts first, the addition order changes. Therefore, IsoMoE passes each token-expert pair in the CPU work list as a separate token with \(k=1\) and a weight of 1.0, so the call returns \(d_j\) itself. Using the merge map, the GPU then merges the \(d_j\) from both sides in routing order with the same FMA and rounding as the CPU (❺ in Fig.~\ref{fig:overview}). Thus, every bit of \(y_t\) stays the same whichever device computes an expert.

\subsection{Lossless 12-bit Experts in CPU Order}
\label{sec:cpu_order}

\begin{figure}[t]
\centering
\includegraphics[width=\columnwidth]{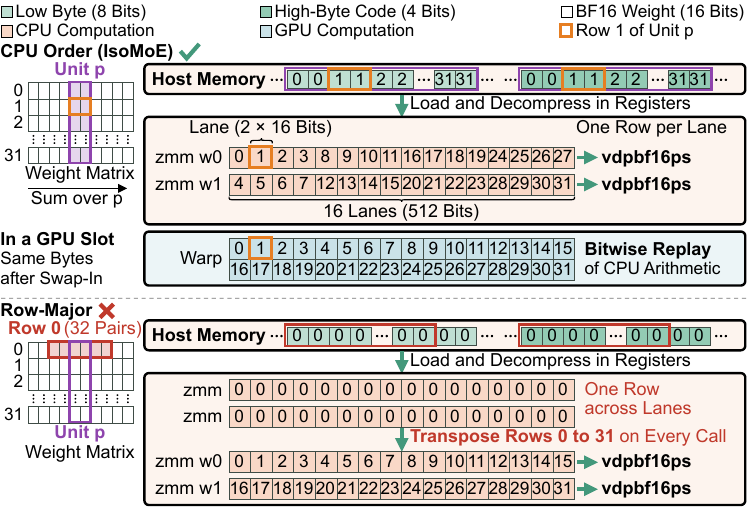}
\caption{12-bit experts in CPU order (top) and row-major order (bottom). Unit~\(p\) holds 32 rows of one BF16 column pair. It is decompressed in registers into \(w_0\) and \(w_1\) for two \texttt{vdpbf16ps}. A GPU warp reads the same bytes.}
\label{fig:cpu_order}
\end{figure}

CPU expert computation is memory-bandwidth-bound (Section~\ref{sec:motivation}). Therefore, IsoMoE adopts the fixed-length lossless 12-bit codewords of Glyd~\cite{koritala2026glyd} to reduce the bytes read by the CPU. The low byte of each BF16 weight is preserved. The high byte is replaced by a 4-bit code, which consists of the sign and a 3-bit offset of the upper 7 exponent bits relative to a base value for each matrix. For a few weights, the offset exceeds the range representable by 3 bits. The original high bytes of the row segments containing these weights (64 weights per segment) are stored in a side table. On average, each weight occupies about 12.1 bits.

The byte arrangement determines what is stored in the registers after decompression. As shown in Fig.~\ref{fig:cpu_order}, unit \(p\) contains 32 rows across a pair of BF16 columns. IsoMoE uses the CPU order shown in the upper half and keeps the low bytes and the codes in two planes, where each unit takes 64 low bytes and 64 codes and the units follow one another. After the CPU expert kernel reads a unit, it decompresses the data in registers (line 8 in Algorithm~\ref{alg:replay}). The \texttt{vpmultishiftqb} instruction extracts the codes, and the \texttt{vpermb} instruction maps the codes to high bytes. The \texttt{vpunpcklbw} and \texttt{vpunpckhbw} instructions interleave the high and low bytes into \(w_0\) and \(w_1\), which are exactly the weight operands for two \texttt{vdpbf16ps} instructions. Each 32-bit lane of these two registers holds a pair of BF16 weights from the same row. The interleaving only occurs within 128-bit blocks. Therefore, after computing a group of 32 rows, the CPU expert kernel uses two \texttt{vpermt2ps} instructions to rearrange the accumulated results in row order (line 11). When an expert is swapped into a GPU slot, these bytes are copied into the slot as is. Each thread of a warp handles one row and replays the CPU arithmetic bitwise.

The lower half shows the default row-major order. A load brings in pairs from a single row, and this row spans all lanes. Therefore, the CPU expert kernel must transpose each group of 32 rows on every call. More memory bandwidth per thread leaves fewer decompression cycles per weight, so the transpose takes up a larger fraction of the cycles. With high bandwidth, CPU expert computation with row-major order becomes even slower than with BF16 experts. With the CPU order, there is no transpose. Even with high bandwidth per thread, the speedup of CPU expert computation approaches the byte ratio (Section~\mbox{\ref{sec:ablation}}).

\subsection{Exact Dual Micro-Batch Pipeline}
\label{sec:pipeline}

\begin{figure}[t]
\centering
\includegraphics[width=\columnwidth]{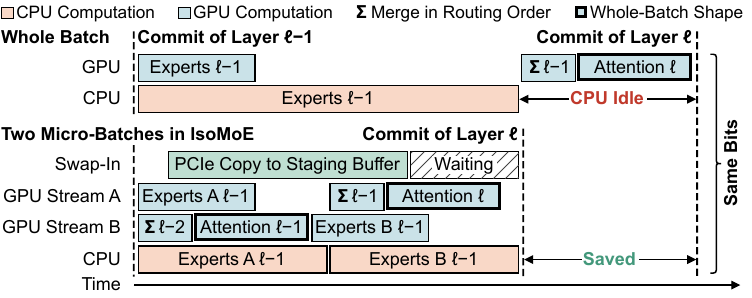}
\caption{Comparison of the whole-batch pipeline and the dual micro-batch pipeline of IsoMoE. Both pipelines produce bitwise identical outputs.}
\label{fig:pipeline}
\end{figure}

IsoMoE decodes with a dual micro-batch pipeline~\cite{liu2024deepseek}. In the whole-batch pipeline shown in the upper half of Fig.~\ref{fig:pipeline}, the GPU waits for the CPU to finish expert computation before it merges (\(\Sigma\)) and computes the attention of the next layer. The CPU idles while the GPU performs these two steps. IsoMoE (lower half) divides a batch into micro-batches A and B, and the two are offset by half a layer. When the CPU computes the experts of A, the GPU merges the expert outputs of B and computes the attention of B, and vice versa. Thus, the CPU no longer idles, and this idle time is saved at every layer.

However, the deterministic modes of SGLang~\cite{sglang2025towards}, vLLM~\cite{bai2026vllmconsistency}, and Megatron-LM~\cite{nvidia2026megatron} do not use a dual micro-batch pipeline. IsoMoE retains the standard GPU libraries that select kernels by shape. Dividing a batch into micro-batches changes the input shapes of the operators, so the low-order bits of the output may change. To solve this problem, IsoMoE computes the non-expert matrix multiplications and attention of a micro-batch using the whole-batch shape. The matrix multiplication input is padded to the whole-batch size with all-zero rows, and each row stays at its position in the whole batch. Attention is scheduled according to the KV partition of the whole batch, and the remaining operators are computed row by row. This ensures that each row in a micro-batch is bitwise identical to the corresponding row of the whole batch.

The timing of swap-ins must not change the output either. A swap-in first copies an expert into the staging buffer over PCIe in the background. At each layer, after the attention and router of micro-batch A are computed, IsoMoE commits the staged swap-ins. It copies them into their slots, updates the residency table, and then splits the experts of this layer. The experts of micro-batch B in this layer are also split after these copies, so the GPU reads the copied bytes for both micro-batches. A swap-in that completes after this point is deferred to the next step for commit. Therefore, an early or late swap-in only affects when the GPU computes that expert.

% ----------------------------------------第五章----------------------------------------
\section{Implementation}
\label{sec:implementation}

We implement IsoMoE on top of KTransformers 0.3.2 \cite{chen2025ktransformers}. We build a library for the CPU expert kernel in C++ with AVX-512 instructions \cite{intel2026sdm}, and KTransformers routes all CPU expert calls to this library. The GPU expert kernel is written in CUDA with inline PTX. The two micro-batches use separate CUDA streams, and a background thread performs swap-ins on its own stream. We convert the released BF16 experts to the 12-bit format offline. IsoMoE runs in exact mode, which follows the design in Section~\ref{sec:design}. For comparison, the GPU expert kernel of a fast mode is implemented in Triton \cite{tillet2019triton}. It uses Tensor Cores to compute matrix multiplications and does not replay the CPU arithmetic. The fast mode does not use the whole-batch shape, and its output changes with placement.

% ----------------------------------------第六章----------------------------------------
\begin{figure*}[t]
\centering
\includegraphics[width=\textwidth]{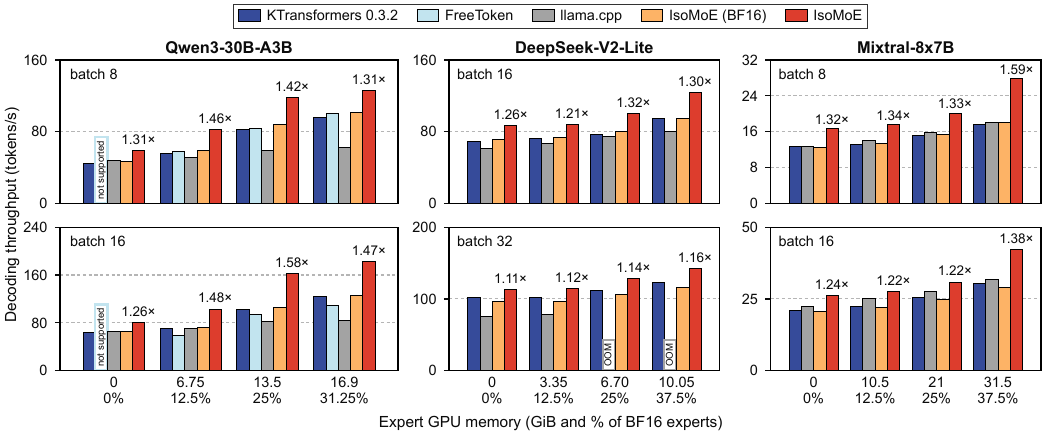}
\caption{Decoding throughput. Labels show speedups of IsoMoE over KTransformers 0.3.2. FreeToken supports only Qwen3 and cannot run at 0 GiB. llama.cpp's 38 GiB KV cache causes OOM at DeepSeek's two largest budgets with batch 32. Qwen3's 24.1 GiB KV cache limits its expert budget to 31.25\%.}
\label{fig:main_performance}
\end{figure*}

\section{Evaluation}

\subsection{Experimental Setup}
\textbf{Platform.} We evaluate IsoMoE on a 48-core Intel Xeon Platinum 8488C CPU \cite{qiu2024m7i} and a 48 GB NVIDIA RTX A6000 GPU \cite{nvidia2022a6000} connected over PCIe 4.0 $\times$16.

\textbf{Models.} We evaluate IsoMoE with three MoE models, namely Qwen3-30B-A3B-Thinking-2507 \cite{yang2025qwen3technicalreport} (Qwen3), DeepSeek-V2-Lite-Chat \cite{deepseekai2024deepseekv2strongeconomicalefficient} (DeepSeek), and Mixtral-8x7B-Instruct \cite{jiang2024mixtralexperts} (Mixtral). We set four expert GPU memory budgets for each model, ranging from 0 to at most 37.5\% of the size of its BF16 experts.

\textbf{Workloads.} Qwen3 answers 82 hard problems from AIME \cite{h4aime2024,zhang2025aime25} and MATH500 \cite{lightman2024let,Hendrycks2021Measuring}. DeepSeek answers 279 MATH500 problems. Mixtral answers 336 math problems, code problems, and chat prompts \cite{lightman2024let,chen2021evaluating,austin2021program,li2023alpacaeval,zheng2023judging}. The batch sizes are 8 and 16 for Qwen3 and Mixtral, and 16 and 32 for DeepSeek.

\textbf{Baselines.} We compare IsoMoE in exact mode with KTransformers 0.3.2 \cite{chen2025ktransformers}, FreeToken \cite{yang2026freetoken}, llama.cpp \cite{gerganov2023llamacpp}, and IsoMoE (BF16). KTransformers 0.3.2 is the base of IsoMoE. It cannot keep individual experts in GPU memory. Therefore, we let it use the GPU expert slots and pipeline of IsoMoE while keeping its own CPU expert kernel. Both systems keep the most frequently selected experts of each layer in GPU memory. We count the selections on prompts outside the evaluation set. FreeToken caches experts in GPU memory with LRU replacement, and llama.cpp keeps the experts of entire layers in GPU memory. IsoMoE (BF16) stores experts in BF16 instead of the 12-bit format and runs in fast mode. We omit HybriMoE \cite{zhong2025hybrimoe} because its released code runs at batch size 1 with 4-bit quantized experts.

\textbf{Metrics.} We measure speed by decoding throughput and placement invariance by whether the outputs are bitwise identical.

\subsection{Placement-Invariance Evaluation}

\begin{table}[t]
\centering
\caption{Output Consistency Across Expert Placements}
\label{tab:placement}
{\footnotesize\setlength{\tabcolsep}{4.7pt}%
\begin{tabular}{@{}lllcc@{}}
\toprule
IsoMoE & Reference & Compared & Identical & Final answers \\
mode & placement & placement & outputs & changed (\%) \\
\midrule
Fast & 20 GPU slots & 41 GPU slots & 1/160 & 8.8 \\
Fast & All-CPU & 41 GPU slots & 0/160 & 10.0 \\
\midrule
Exact & 20 GPU slots & 41 GPU slots & 160/160 & 0 \\
Exact & All-CPU & 41 GPU slots & 160/160 & 0 \\
Exact & 41 GPU slots & Online & 56/56 & 0 \\
\bottomrule
\end{tabular}}
\end{table}

Table~\ref{tab:placement} shows that the exact mode of IsoMoE makes the outputs bitwise identical under different expert placements. We let Qwen3 solve 160 problems from AIME \cite{h4aime2024,zhang2025aime25} and MATH500 \cite{lightman2024let,Hendrycks2021Measuring} with a batch size of 8 and greedy decoding. Each request generates at most 16K tokens. The all-CPU placement keeps all experts in host memory, and the CPU computes them. The 20-slot and 41-slot placements are static. They keep the 20 or 41 most frequently selected experts of each layer in GPU expert slots, and the GPU computes these experts. In fast mode, moving from 20 slots or all-CPU to 41 slots changes at least 159 of the 160 outputs. The final answer given after the reasoning also changes for 8.8\% to 10.0\% of the problems. In exact mode, all outputs stay bitwise identical under the same changes. IsoMoE runs the whole-batch pipeline under the all-CPU placement and the dual micro-batch pipeline under the 20-slot and 41-slot placements. Thus, the two pipelines also produce identical outputs. With online placement, IsoMoE swaps the experts that each layer selected most often in recent decode steps into the 41 GPU expert slots of that layer. On 56 problems that span the range of output lengths, 49,712 swap-ins leave the outputs bitwise identical to those under the 41-slot static placement. Therefore, new swap-in strategies can be verified through bitwise comparison against static placement.

IsoMoE's outputs are also bitwise identical to those of the original KTransformers 0.3.2 \cite{chen2025ktransformers}, which computes all experts on the CPU. On Qwen3, both use BF16 for non-expert layers and do not divide the batch into micro-batches. On 16 problems, we raise the output cap to 32K tokens. All outputs of IsoMoE with 41 slots match those of the original, and the longest output has 32,264 tokens. On DeepSeek, the batch sizes are 16 and 32. To compare the computation step by step on long sequences, we fix the output length of each request at 4,096 tokens. With 10, 21, and 31 slots, IsoMoE matches the original in the logits and per-layer output hashes at every step. On Mixtral, the batch sizes are 8 and 16, and each request generates 256 tokens. With 1, 2, and 4 slots, IsoMoE also matches the original in the logits and per-layer output hashes at every step. When both modes use the same 12-bit experts, the decoding throughput of exact mode is 1.02\(\times\) that of fast mode on Qwen3 with 41 slots and 0.97\(\times\) on DeepSeek with 21 slots. Both are within the variation between repeated runs, so exact mode incurs no measurable speed loss.

\subsection{End-to-End Performance}

As shown in Fig.~\ref{fig:main_performance}, IsoMoE achieves the highest decoding throughput across all evaluated expert GPU memory budgets and both batch sizes for the three models. Compared with the fastest baseline in each setting, it achieves an average speedup of 1.26\(\times\) and a maximum speedup of 1.54\(\times\).

With all experts on the CPU, the throughput of IsoMoE is 1.16 to 1.34 times that of IsoMoE (BF16), because the CPU reads about one quarter fewer expert weight bytes. With some experts on the GPU, the 12-bit format also allows the same GPU memory to hold more experts, enabling the GPU to compute more selected experts. For example, at 13.5 GiB on Qwen3 with a batch size of 16, IsoMoE (BF16) is the fastest baseline, and IsoMoE reaches 1.54 times the throughput of this baseline.

The baselines use different arithmetic to compute experts on the CPU and GPU, so they cannot guarantee unchanged outputs after moving experts. To guarantee unchanged outputs, they must keep all experts on the CPU. IsoMoE keeps its outputs unchanged under every expert GPU memory budget. At the maximum budget of each model, the throughput of IsoMoE is 1.39 to 2.76 times that of the fastest baseline with all experts on the CPU.

\subsection{Ablation Study}
\label{sec:ablation}

\begin{table}[t]
\centering
\caption{Ablation on Qwen3 at batch~8 and a 13.5~\textnormal{GiB} expert GPU memory budget. The baseline is IsoMoE (BF16). \textnormal{(a)}~12-bit format. \textnormal{(b)}~More GPU experts. \textnormal{(c)}~Exact mode.}
\label{tab:ablation}
{\footnotesize\setlength{\tabcolsep}{4pt}\renewcommand{\arraystretch}{1.12}%
\begin{tabular}{@{}lccccc@{}}
\toprule
\raisebox{4pt}{Configuration} & \shortstack{GPU\\slots} & \shortstack{GPU memory\\(GiB)} & \shortstack{Hit rate\\(\%)} & \shortstack{Throughput\\(tokens/s)} & \shortstack{Speedup\\over baseline} \\
\midrule
Baseline & 32 & 13.5 & 67 & 87.7 & 1.00$\times$ \\
+(a) & 32 & 10.3 & 67 & 105.1 & 1.20$\times$ \\
+(a)+(b) & 41 & 13.2 & 75 & 118.5 & 1.35$\times$ \\
+(a)+(b)+(c) & 41 & 13.2 & 75 & 117.9 & 1.34$\times$ \\
\bottomrule
\end{tabular}}%
\end{table}

Table~\ref{tab:ablation} breaks down the benefits of format compression and increased GPU expert capacity on Qwen3 with a batch size of 8 and an expert GPU memory budget of 13.5 GiB. With 32 GPU expert slots fixed per layer, the 12-bit format (a) increases decoding throughput to 1.20 times that of IsoMoE (BF16) and reduces the GPU memory reserved for expert slots from 13.5 GiB to 10.3 GiB. Using the saved memory to increase GPU expert capacity (b) raises the number of slots per layer to 41 and the GPU expert hit rate from 67\% to 75\%, which further improves throughput by 13\% over (a). With exact mode (c) enabled, the full IsoMoE reaches a throughput of 117.9 tokens/s, which is 1.34 times that of the baseline.

We also compare two byte arrangements of the same 12-bit codewords in CPU expert computation on Qwen3. We vary the CPU thread count from 1 to 40. Across these configurations, BF16 expert computation reaches an effective bandwidth of about 3.3 to 15 GB/s per thread, while expert computation in CPU order remains 1.28 to 1.35 times as fast as BF16, close to the byte ratio. With row-major order, the speed ratio is 1.02 to 1.05 at about 8 GB/s per thread and drops to 0.68 to 0.70 at about 15 GB/s. Since both arrangements have the same compressed data size, this comparison shows that the CPU order preserves the compression benefit at higher memory bandwidth per thread.

\section{Conclusion}

In this paper, we present IsoMoE, an efficient placement-invariant hybrid MoE inference system. IsoMoE bitwise replays the CPU arithmetic on the GPU and stores lossless 12-bit experts in CPU order, enabling fast CPU decompression. On three MoE models, IsoMoE produces bitwise identical outputs across different expert placements and decodes faster than the fastest baseline in every setting, with an average speedup of 1.26\(\times\) and a maximum speedup of 1.54\(\times\). In IsoMoE, changing expert placement only affects inference speed and does not change the outputs.

% \section*{Acknowledgment}

% The preferred spelling of the word ``acknowledgment'' in America is without 
% an ``e'' after the ``g''. Avoid the stilted expression ``one of us (R. B. 
% G.) thanks $\ldots$''. Instead, try ``R. B. G. thanks$\ldots$''. Put sponsor 
% acknowledgments in the unnumbered footnote on the first page.

% \clearpage
\bibliographystyle{IEEEtran}
\bibliography{ref}

\end{document}